# Mode Multigrid - A novel convergence acceleration method

Yilang Liu, Weiwei Zhang[*], Jiaqing Kou

*( School of Aeronautics, Northwestern Polytechnical University, No. 127 Youyi West Road, Xi'an 710072, China )*

**Abstract:** This paper proposes a mode multigrid (MMG) method, and applies it to accelerate the convergence of the steady state flow on unstructured grids. The dynamic mode decomposition (DMD) technique is used to analyze the convergence process of steady flow field according to the solution vectors from the previous time steps. Unlike the traditional multigrid method, we project the flowfield solutions from the physical space into the modal space, and truncate all the high-frequency modes but only the first-order mode are retained based on the DMD analysis. The real solutions in the physical space can be obtained simply by the inverse transformation from the modal space. The developed MMG method ingeniously avoids the complicated process of coarsening computational mesh, and does not need to make any change for the grid in physical space. Therefore, it is very convenient to be applied to any numerical schemes with just little change for the flow solver, which is also suitable for unstructured grids and easy for parallel computing. Several typical test cases have been used to verify the effectiveness of the proposed method, which demonstrates that the MMG can dramatically reduce the number of iterative steps for the different mesh types, different accuracy of spatial discretization and different time-marching schemes. The method is 3 to 6 times faster than the original method while ensuring the computational accuracy.

## 1. Introduction

Computational fluid dynamics (CFD) researchers have kept pursuing the accuracy and the efficiency of numerical simulations. With the development of engineering, large amounts of meshes need to be used to adapt the more and more complex configurations. Therefore, how to accelerate the convergence for flow fields and improve the computational efficiency are still very important for CFD. Various methodologies have been used for accelerating the convergence to steady state, which can be mainly divided into three types. The first one is developing high-efficiency time-marching methods, such as the implicit time-marching schemes. The second type is adopting some acceleration techniques, such as the local time-stepping [1], enthalpy damping [2], residual smoothing [3] and the multigrid method [4]. The third one is the vector extrapolation method.

For the time-marching method, the explicit scheme is firstly used for flow solvers, and the most popular and widespread one is the Runge-Kutta scheme [5~6]. It has many advantages, such as low memory cost, high computational accuracy and easy for paralleling. However, the maximum time step for the explicit scheme is dependent on the cell size, since it should fulfill the CFL condition to keep stable. The computational efficiency will be extremely reduced when the grid scale is very small. Thus, the implicit time-marching schemes are proposed to accelerate the convergence, such as the implicit Symmetric Gauss-Seidel (SGS) method [7~9], Lower-Upper Symmetric Gauss-Seidel (LUSGS) method [10~11], Alternating Direction Implicit (ADI) method [12] and Generalized Minimal Residual (GMRES) method [13~14]. Compared with the explicit scheme, the implicit time-marching method can adopt much larger time step, which has high computational efficiency, especially for the turbulence flow in high Reynolds number.

For the second acceleration method, one of the most typical techniques is the multigrid (MG) method [4, 15~30]. The multigrid methodology is a very powerful acceleration technique, which mainly contains two realizable ways - geometric multigrid (h-multigrid) and polynomial multigrid (p-multigrid) method. The basic idea of the geometric multigrid is to solve the governing equations on a series of successively coarser grids in order to reduce the different frequency components of the solution errors. It is relatively convenient to be used on structured grids since the ordered storage of the grid information makes it easy for generating the coarse grids. Mulder [18] and Pierce [19] proposed the semi-coarsening (or J-coarsening) method in order to apply it on the anisotropic structured grid, which has been widely used for the viscous flows on the highly stretched meshes [20~21]. As compared to the structured grid, the geometric multigrid technique is quite difficult to implement on the unstructured meshes, because it is much more involved to construct a series successively coarsened meshes from the initial grid with no particular ordering. Wesseling [22] pointed out that it is difficult to define a sequence of coarser grids starting from an irregular fine grid, and the coarse grids may insufficiently to support the real geometry, especially for the strongly curved boundaries. Mavriplis [23] considered that one of the important drawbacks of multigrid technique on unstructured meshes is that storing the different coarse levels will dramatically increase the computational memory, particularly in 3D. For the high-order schemes, such as discontinuous Galerkin (DG), the numerical integration might lead to excessive matrix assembly costs as a consequence of the lack of efficient quadrature rules for agglomerated elements [24~25]. Thus, it is more expensive to apply the geometric multigrid method for the high-order schemes. Moreover, there are also many other difficulties for the geometric multigrid to be consideration, such as the proper data transfer between successive meshes, the treatment for the highly stretched anisotropic hybrid grids and the paralleling computing [26~27]. Different from the geometric multigrid, p-multigrid method [28~30] solve the discrete equations by recursively iterating on solution approximations of different polynomial orders instead of different grids. However, it is technically difficult to develop high-order schemes on unstructured grids, and developing p-multigrid method is also complex. Due to this reason, the implementation of multigrid method on unstructured grids is limited as well.

The vector extrapolation method, which is originally used to accelerate solving the system of linear algebraic equations, becomes a new research hotspot in recent years. Two of the most widely used are the minimal polynomial extrapolation (MPE) method, proposed by Cabay & Jackson [31], and the reduce rank extrapolation (RRE) method, proposed by Mešina [32]. Sidi [33] has well reviewed of the two methods. Subsequently, CFD researchers introduce the vector extrapolation method to accelerating the convergence of flow solvers. The essential idea of the method is that an improved initial guess for the iterative process can be obtained according to the polynomial interpolation or feature extraction techniques from some previous time steps. Intuitively, the acceleration method is easier to be comprehended for unsteady flow simulation. In the dual time-stepping schemes, the initial flow field for the physical time step adopts the improved guess values according to extraction method by using several previous flow fields of the physical time, instead of the flow field of just one former step. The modification can reduce the number of pseudo-time iteration steps to a certain extent. Gong [34] proposed to adopt the Taylor series polynomial interpolation methodology to improve the dual time-stepping scheme. The method has been used to solve several subsonic unsteady periodic flow fields, and also researched the effects of different interpolation schemes. The computational efficiency has been improved over 50% compared with the initial scheme. More recently, Liu [35] also accelerated the dual time-stepping scheme by adopting the dynamic mode decomposition (DMD) technique. The DMD modes are extracted and interpolated by choosing reasonable flow fields in previous physical time steps, which can also improve the computational efficiency. For accelerating the steady flow solver, Markovinovi [36] used a proper orthogonal decomposition (POD) method to determine an improved guess for the iterative initial flow field based on the information of the previous time steps. Andersson [37] adopted the similar approach by using DMD technique, and applied it for simulating the two-dimensional cascade flow, which reduced the number of iterations for about 30% compared with the original method. In addition, Kaminsky [38] and Djeddi [39] also used the POD technique to improve the convergence for the explicit time-stepping schemes. However, the effect of accelerating convergence for the steady flow is not yet perfect by using the vector extrapolation method, and the mechanism of acceleration for the method is not very clear at present. As mentioned in [39], "the solution snapshots require a careful selection process. The effectiveness and robustness of the acceleration technique are directly related to the amount of information that the collected snapshots provide, and selecting the best snapshots is more art than science".

After the deep research of the various acceleration methods, this paper combines the advantages of multigrid method with the vector extrapolation method, and develops a novel mode multigrid method (MMG) based on the DMD technique, and applies it to accelerate the steady state flow solver. The proposed MMG method can filter out different frequency components of the solution errors, and reduce the number of iterations effectively and robustly, which can achieve the similar effect of the traditional multigrid method. On the other hand, it also properly maintains the advantages of the vector extrapolation method, simple to be implemented and universal to be applied.

The outline of this paper is as follows. Section 2 describes the CFD governing equations and the numerical method briefly. The developed MMG method is introduced in detail in Section 3. Section 4 presents several numerical examples to verify the performance of the proposed acceleration method, and the conclusions are drawn in Section 5.

## 2. Governing equations and the numerical method

The integral form of three-dimensional Navier-Stocks equations can be written as:

$$\frac{\partial}{\partial t}\iint_\Omega \boldsymbol{Q} d\Omega + \oint_{\partial\Omega} \boldsymbol{F}(\boldsymbol{Q})\cdot\boldsymbol{n} d\Gamma = \oint_{\partial\Omega} \boldsymbol{G}(\boldsymbol{Q})\cdot\boldsymbol{n} d\Gamma \tag{1}$$

where $\Omega$ is the control volume; $\partial\Omega$ is the boundary of control volume; and $\boldsymbol{n}=(n_x,n_y,n_z)^T$ denotes the unit outward normal vector to the boundary. The vector of conservative variables $\boldsymbol{Q}$, inviscid fluxes $\boldsymbol{F}(\boldsymbol{Q})=(\boldsymbol{F}_x(\boldsymbol{Q}),\boldsymbol{F}_y(\boldsymbol{Q}),\boldsymbol{F}_z(\boldsymbol{Q}))$ and viscous fluxes $\boldsymbol{G}(\boldsymbol{Q})=(\boldsymbol{G}_x(\boldsymbol{Q}),\boldsymbol{G}_y(\boldsymbol{Q}),\boldsymbol{G}_z(\boldsymbol{Q}))$ are given as follows:

$$\boldsymbol{Q}=\begin{Bmatrix}\rho\\\rho u\\\rho v\\\rho w\\E\end{Bmatrix} \boldsymbol{F}_x(\boldsymbol{Q})=\begin{Bmatrix}\rho u\\\rho u^2+p\\\rho uv\\\rho uw\\u(E+p)\end{Bmatrix} \boldsymbol{F}_y(\boldsymbol{Q})=\begin{Bmatrix}\rho v\\\rho uv\\\rho v^2+p\\\rho vw\\v(E+p)\end{Bmatrix} \boldsymbol{F}_z(\boldsymbol{Q})=\begin{Bmatrix}\rho w\\\rho uw\\\rho vw\\\rho w^2+p\\w(E+p)\end{Bmatrix}$$

$$\boldsymbol{G}_x(\boldsymbol{Q})=\left\{0\quad \tau_{xx}\quad \tau_{xy}\quad \tau_{xz}\quad u\tau_{xx}+v\tau_{xy}+w\tau_{xz}+\frac{1}{(\gamma-1)\Pr}\frac{\partial T}{\partial x}\right\}^T$$

$$G_y(Q) = \left\{0 \quad \tau_{xy} \quad \tau_{yy} \quad \tau_{yz} \quad u\tau_{xy} + v\tau_{yy} + w\tau_{yz} + \frac{1}{(\gamma-1)\Pr}\frac{\partial T}{\partial y}\right\}^T$$

$$G_z(Q) = \left\{0 \quad \tau_{xz} \quad \tau_{yz} \quad \tau_{zz} \quad u\tau_{xz} + v\tau_{yz} + w\tau_{zz} + \frac{1}{(\gamma-1)\Pr}\frac{\partial T}{\partial z}\right\}^T$$

where the viscous stress are

$$\tau_{xx} = 2\mu\frac{\partial u}{\partial x} - \frac{2}{3}\mu(\frac{\partial u}{\partial x} + \frac{\partial v}{\partial y} + \frac{\partial w}{\partial z}), \quad \tau_{xy} = \mu(\frac{\partial u}{\partial y} + \frac{\partial v}{\partial x})$$

$$\tau_{yy} = 2\mu\frac{\partial v}{\partial y} - \frac{2}{3}\mu(\frac{\partial u}{\partial x} + \frac{\partial v}{\partial y} + \frac{\partial w}{\partial z}), \quad \tau_{xz} = \mu(\frac{\partial u}{\partial z} + \frac{\partial w}{\partial x})$$

$$\tau_{zz} = 2\mu\frac{\partial w}{\partial z} - \frac{2}{3}\mu(\frac{\partial u}{\partial x} + \frac{\partial v}{\partial y} + \frac{\partial w}{\partial z}), \quad \tau_{yz} = \mu(\frac{\partial v}{\partial z} + \frac{\partial w}{\partial y})$$

where $\rho$ denotes the density; $u$, $v$ and $w$ are the $x$, $y$ and $z$ direction components of the velocity vector; $p$ is the pressure; $E$ is the total energy per unite volume; $\mu$ is the dynamic molecular viscosity; $T$ is the temperature; $\Pr$ is the Prandtl number; and $\gamma$ is the ratio of specific heats. For the ideal gas $\gamma$ is equal to 1.4. According to Sutherland's law, the dynamic viscosity is given by

$$\mu = \mu_{ref}\frac{T_{ref} + S_0}{T + S_0}(\frac{T}{T_{ref}})^{\frac{3}{2}} \tag{2}$$

where $T_{ref}$ and $\mu_{ref}$ are physical constants of reference temperature and viscosity, and $S_0$ is the Sutherland temperature. The values of them are $T_{ref} = 273.15K$, $\mu_{ref} = 1.716 \times 10^{-5} \, kg/(m \cdot s)$ and $S_0 = 110K$, respectively. The equations of state for the ideal gas is

$$p = (\gamma - 1)[E - \frac{\rho}{2}(u^2 + v^2 + w^2)] \tag{3}$$

In the cell-centered finite volume method, the computational domain is divided into non-overlapping control volumes that completely cover the domain. The interface variables are derived from the average values of the grid cells to calculate the fluxes of control volumes. Through spatial discretization, the equations of the integral form are translated to ordinary differential equations in time, and the flow variables are obtained by the time marching method. The semi-discrete finite-volume formulation of the flow equations is

$$\frac{d\overline{\boldsymbol{Q}}_i}{dt} = -\frac{1}{|\Omega_i|} \sum_{m \in N(i)} \sum_{j=1}^{q} |\Gamma_{i,m}| \omega_j (\boldsymbol{F}(\boldsymbol{Q}(x_j, y_j, z_j)) - \boldsymbol{G}(\boldsymbol{Q}(x_j, y_j, z_j))) \cdot \boldsymbol{n}_{i,m} \quad (4)$$

where $|\Omega_i|$ denotes the volume for the $i$ th cell; $N(i)$ is the set of cells neighboring the cell $i$; and $\Gamma_{i,m}$ is the interface area between cell $i$ and the neighbor cell $m$. $q$ and $\omega_j$ denote the Gauss integral points and the weight coefficients of the interface, respectively; and $\boldsymbol{n}_{i,m}$ is the outer normal vector of the interface. The average conservative vector of the control volume $i$ is $\overline{\boldsymbol{Q}}_i$, which is computed by

$$\overline{\boldsymbol{Q}}_i = \frac{1}{|\Omega_i|} \int_{\Omega_i} \boldsymbol{Q}(x, y, z) d\Omega \quad (5)$$

The numerical fluxes at the right hand of equation (4) can be evaluated by upwind schemes. According to the Godunov-type method, the interface normal fluxes are calculated by the Riemann solver:

$$\boldsymbol{F}(\boldsymbol{Q}_{i,m}) \cdot \boldsymbol{n}_{i,m} \approx \tilde{\boldsymbol{F}}(\boldsymbol{Q}_{i,m}^L, \boldsymbol{Q}_{i,m}^R, \boldsymbol{n}_{i,m}) \quad (6)$$

where the superscript "L" and "R" denote the states of flow variables approaching to the left and right sides of the cell interface, respectively. This paper adopts the Roe [40] scheme to compute the numerical fluxes, where $\boldsymbol{Q}_{i,m}^L$ and $\boldsymbol{Q}_{i,m}^R$ are used to evaluate the Roe's average states.

The semi-discrete formulation of flow equation (4) can be translated to ordinary differential equations in time after obtaining the discrete numerical fluxes:

$$\frac{d\overline{\boldsymbol{Q}}_i}{dt} = \overline{R}_i \quad (7)$$

where $\overline{R}_i$ denotes the sum of inviscid and viscid fluxes. Finally, the semi-discrete system (7) can be marched in time using the explicit or implicit schemes. In this paper, both the explicit Runge-Kutta method [5] and the implicit symmetric Gauss-Seidel (SGS) method [7] are implemented for the time marching.

**3. The mode multigrid method based on the DMD technique**

The convergence of the steady flow field can be considered as the process that solution errors generated from the wall boundary propagate to the far field until it is eliminated. Both the high-frequency and low-frequency components of errors can be produced during the flow solving. And the basic idea of the traditional multigrid method is to solve the governing equations on a hierarchy of grids in order to damp different frequency errors. However, the traditional multigrid is much more complex when implemented on the unstructured grids. Due to the random storage of grid information, it is not convenient for parallel computing and has not been widely used in applications. We propose a new ingenious way to solve the problem.

### 3.1 The DMD technique

The DMD technique is a data-driven approach to extract the coherent flow structures and analyze the dominant flow dynamics, which was introduce by Schmid[41]. It has been widely applied to the flow mechanics, such as turbulent boundary layer interaction [42~43], flow transition [44~45], the backward-facing step flow [46] and the vortex-induced vibrations [47~48]. More details have been reviewed in Reference [49]. The DMD method can be described by singular value decomposition (SVD) of the snapshot matrix. The snapshot sequence with $N$ samples can be obtained from the flow field solutions in the previous time steps, which is described as $\{\mathbf{x}_1, \mathbf{x}_2, \mathbf{x}_3, ..., \mathbf{x}_N\}$, where the $i$th snapshot is $\mathbf{x}_i \in \mathbb{C}^M$ (N << M). The conservative variables $(\rho, \rho u, \rho v, E)$ are selected as the snapshot. We assume a linear dynamical system for mapping the current flow field to the subsequent flow field:

$$\mathbf{x}_{i+1} = \mathbf{A}\mathbf{x}_i \tag{8}$$

where $\mathbf{A} \in \mathbb{C}^{M \times M}$ is the system matrix containing a particularly large number of entries. Because the linear relationship is assumed, the dynamical characteristics are contained in the eigenvalues of matrix $\mathbf{A}$. In order to obtain dominant eigenvalues accurately, the order of the high-dimensional system matrix $\mathbf{A}$ should be reduced. We then form two matrices:

$$\mathbf{X} = [\mathbf{x}_1, \mathbf{x}_2, \mathbf{x}_3, ..., \mathbf{x}_{N-1}] \tag{9}$$

$$\mathbf{Y} = [\mathbf{x}_2, \mathbf{x}_3, \mathbf{x}_4, ..., \mathbf{x}_N] \tag{10}$$

Using the linear process in (8), a matrix constructed as a Krylov sequence is obtained:

$$\mathbf{Y} = [\mathbf{x}_2, \mathbf{x}_3, \mathbf{x}_4, ..., \mathbf{x}_N] = [\mathbf{A}\mathbf{x}_1, \mathbf{A}\mathbf{x}_2, \mathbf{A}\mathbf{x}_3, ..., \mathbf{A}\mathbf{x}_{N-1}] = \mathbf{A}\mathbf{X} \tag{11}$$

Then the DMD is achieved by a similarity transformation of the system matrix, and a similar matrix $\widetilde{\mathbf{A}}$ should be constructed to replace the full-order matrix $\mathbf{A}$. Firstly, we seek an invertible matrix by performing SVD on the snapshot matrix $\mathbf{X}$:

$$\mathbf{X} = \mathbf{U}\mathbf{\Sigma}\mathbf{V}^H \tag{12}$$

$$\mathbf{A} = \mathbf{U}\widetilde{\mathbf{A}}\mathbf{U}^H \tag{13}$$

where $\mathbf{\Sigma}$ contains $r$ non-zero singular values $\{\sigma_1,...,\sigma_r\}$ in its diagonal. From (12), we have $\mathbf{U}^H\mathbf{U} = \mathbf{I}, \mathbf{U} \in \mathbb{C}^{M \times r}$ and $\mathbf{V}^H\mathbf{V} = \mathbf{I}, \mathbf{V} \in \mathbb{C}^{r \times N}$. Matrix $\widetilde{\mathbf{A}}$ can be calculated by minimizing the Frobenius norm of the difference between $\mathbf{Y}$ and $\mathbf{AX}$:

$$\underset{\mathbf{A}}{\text{minimize}} \|\mathbf{Y} - \mathbf{AX}\|_F^2 \tag{14}$$

From (12) and (13), (14) is expressed as:

$$\underset{\widetilde{\mathbf{A}}}{\text{minimize}} \|\mathbf{Y} - \mathbf{U}\widetilde{\mathbf{A}}\mathbf{\Sigma}\mathbf{V}^H\|_F^2 \tag{15}$$

$\mathbf{A}$ is then approximated by $\widetilde{\mathbf{A}}$:

$$\mathbf{A} \approx \widetilde{\mathbf{A}} = \mathbf{U}^H\mathbf{Y}\mathbf{V}\mathbf{\Sigma}^{-1} \tag{16}$$

Because $\widetilde{\mathbf{A}}$ is the similar matrix of $\mathbf{A}$, eigenvalues of $\widetilde{\mathbf{A}}$ are some main eigenvalues of $\mathbf{A}$. $\widetilde{\mathbf{A}}$ has the eigenvalue $\mu_j$ which makes $\widetilde{\mathbf{A}}\mathbf{w}_j = \mu_j\mathbf{w}_j$, where $\mathbf{w}_j$ is the eigenvector of the $j$th eigenvalue. Thus, the system matrix $\widetilde{\mathbf{A}}$ and the DMD mode $\mathbf{\Phi}$ can be expressed as

$$\begin{aligned}\widetilde{\mathbf{A}} = \mathbf{W}\mathbf{N}\mathbf{W}^{-1}, \quad \mathbf{N} = diag(\mu_1,...,\mu_r) \\ \mathbf{\Phi} = \mathbf{U}\mathbf{W}\end{aligned} \tag{17}$$

where the matrix $\mathbf{W}$ contains each eigenvector of $\widetilde{\mathbf{A}}$ in its column. And the Mode amplitude vector $\boldsymbol{\alpha}$ is represented by:

$$\boldsymbol{\alpha} = \mathbf{W}^{-1}\mathbf{z}_1 = \mathbf{W}^{-1}\mathbf{U}^H\mathbf{x}_1, \quad \boldsymbol{\alpha} = [\alpha_1,...,\alpha_r]^T \tag{18}$$

where $\alpha_i$ denotes the amplitude of the $i$th mode, which represents the modal contribution on the initial snapshot $\mathbf{x}_1$. DMD modes are ordered by their amplitudes (entries of vector $\boldsymbol{\alpha}$). Substitute (13), (17) and (18) into (8), the flow field at any time instant is given by:

$$\mathbf{x}_i = \mathbf{\Phi}\mathbf{\Lambda}^{i-1}\boldsymbol{\alpha} = \sum_{j=1}^{r}\mathbf{\Phi}_j(\mu_j)^{i-1}\alpha_j \tag{19}$$

Then, the Matrix $\mathbf{Y}$ can be expressed as:

$$Y = [\mathbf{x}_2, \mathbf{x}_3, ..., \mathbf{x}_N] = \mathbf{\Phi D}_\alpha \mathbf{V}_{and} = [\mathbf{\Phi}_1, \mathbf{\Phi}_2, ..., \mathbf{\Phi}_r] \begin{bmatrix} \alpha_1 & & & \\ & \alpha_2 & & \\ & & . & \\ & & & \alpha_r \end{bmatrix} \begin{bmatrix} \mu_1 & (\mu_1)^2 & . & (\mu_1)^{N-1} \\ \mu_2 & (\mu_2)^2 & . & (\mu_2)^{N-1} \\ . & . & . & . \\ \mu_r & (\mu_r)^2 & . & (\mu_r)^{N-1} \end{bmatrix} \quad (20)$$

In (20), the flow evolution is represented by the Vandermonde matrix $\mathbf{V}_{and}$, which contains *r* eigenvalues of the matrix $\mathbf{A}$. Then, the flow field can be reconstructed in the modal space by using equation (20).

**3.2 Dynamic mode analysis for the convergence of steady state flow**

In this section, we analyze the dynamic modes for the steady state flow in detail by using the numerical case of subsonic inviscid flow past a NACA0012 airfoil. The free stream Mach number is $Ma = 0.63$ and the angle of attack is $\alpha = 2°$. For the entire convergence process of the steady state flow, each snapshot is taken every 10 pseudo time steps from the beginning of the iteration, and we perform the mode analysis successively by taking every 100 snapshots. Figure 1 shows the evolution of the modal coefficients for the first 10 modes with DMD and the first 5 modes with POD respectively. The results demonstrate that the first modal coefficient approaches to a constant, and all of other modal coefficients converge to zero for both method. A single mode is obtained for POD, and the complex conjugate modes are obtained except for the first mode by using DMD. It can be clearly seen from the process of the modal coefficients, the convergence of steady state flow can be considered in the perspective of modal space, that is, the first-order modal coefficient of the flow modes approaches to a constant, and all of other-order modal coefficients decay to zero. There are also some differences for the first order mode coefficient. The POD behaves as an oscillatory convergence, whereas the DMD nearly maintain unchanged during the entire flow revolutions. It indicates that the DMD has the ability to predict the stable state for the oscillating data information. Thus, we specifically focus on the second time of the mode analysis, which are the 1000 to 2000 pseudo time steps. And the DMD flow modes for the first 10 orders are displayed in Figure 2. The frequency spectrum with Fourier analysis for the modal coefficients of both POD and DMD are shown in Figure 3. The results show that although both methods can extract the main information of the flow feature, there are still distinctions. The POD decomposition of the flow structures is based on the kinetic energy content, whereas the DMD is based on the temporal frequency of oscillation. Therefore, each POD mode may contain different frequency components of the flow characteristics, and the single frequency is obtained for each DMD mode. Compared with the POD modes, DMD can separate different frequency components more clearly. This is very convenient to filter out the high-frequency perturbations of the flow field in the modal space.

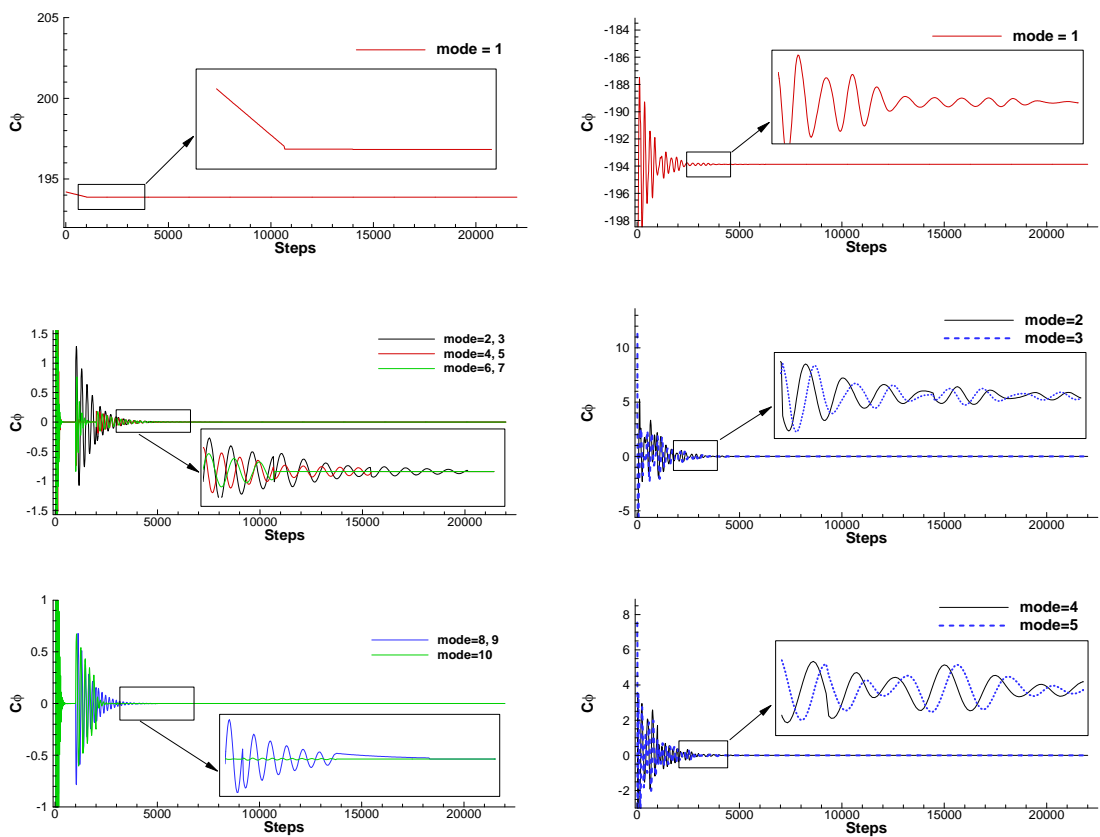

Figure 1 The modal coefficients for the steady state flow. Left: DMD mode; Right: POD mode.

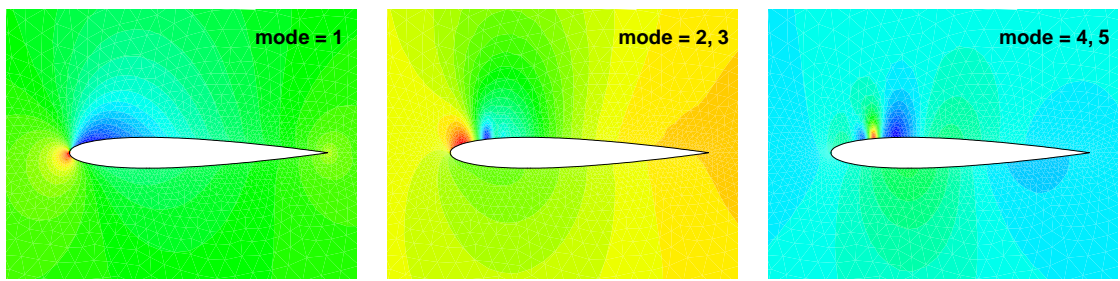

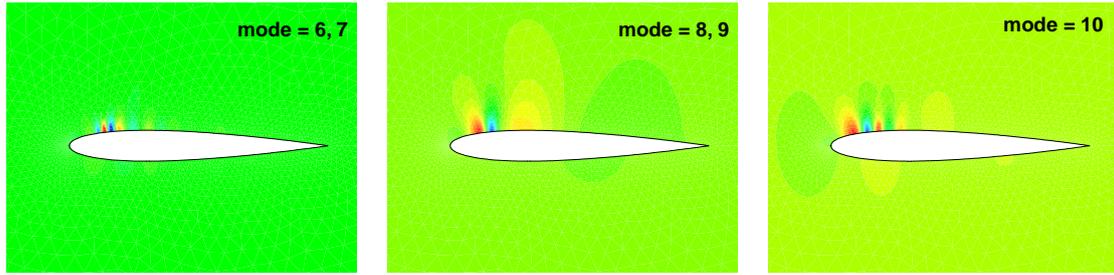

Figure 2 The DMD flow modes for the 1000~2000 iterative steps

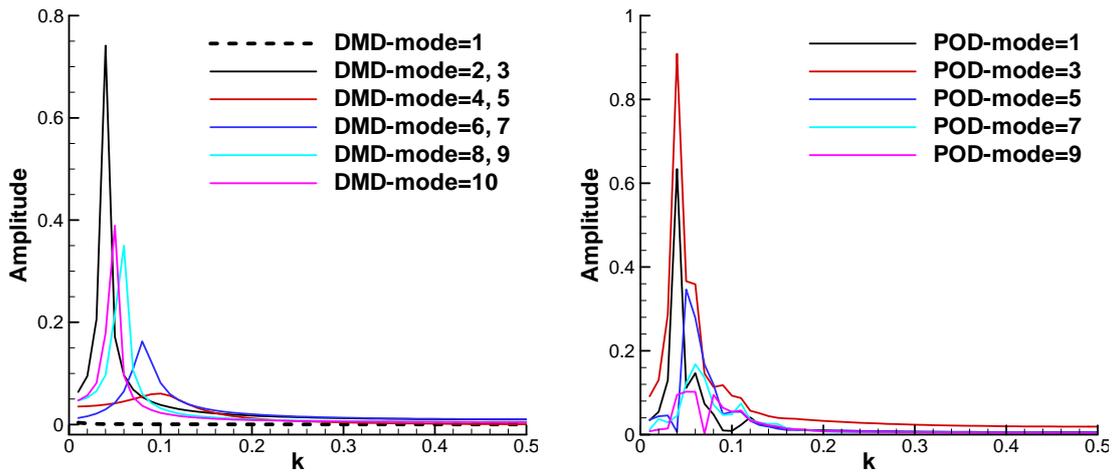

Figure 3 The frequency spectrum with Fourier analysis for modal coefficients of the POD and DMD

Figure 4 shows the comparisons of the first-order DMD mode obtained by the first (1~1000 time steps), second (1000~2000 time steps) and the final times of the DMD analysis respectively. It can be seen from the results, although all of the DMD modes are time-varying, the first-order mode is relatively stable with respect to other higher modes. The stable mode changes little during the whole iterations of the flow field, and the distributions of the mode obtained by the second DMD analysis have been very similar to the final convergence ones. The modal coefficients in Figure 1 also demonstrate that the DMD can rapidly and accurately predict the first-order stable mode only by a few previous iterative steps. Figure 5 displays the comparisons of the first-order DMD mode for the final time with the distributions of the convergence flow field. The two different flow fields are equally divided into 30 contour lines according to the maximum and minimum values. As can be seen from the results, the contour lines can completely coincide with each other. Therefore, we deem that the first-order stable mode obtained by the DMD represents the convergence solutions for the steady state flow, and only a proportion factor is different.

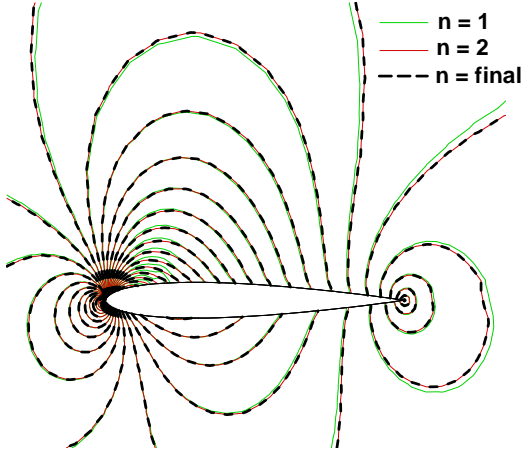 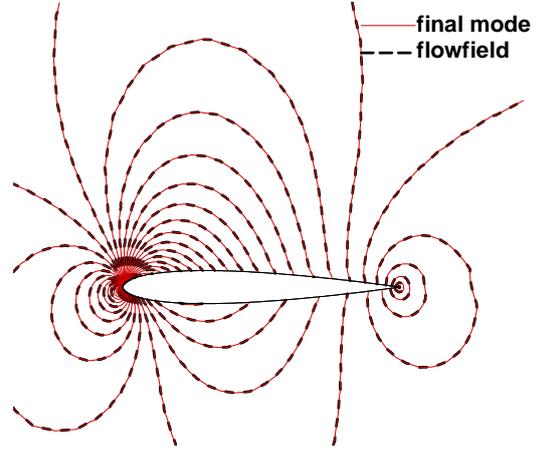

Figure 4 The comparisons of the first-order mode for the first, second and the final times of the DMD analysis

Figure 5 The comparisons of the first-order DMD flow mode for the final time with the convergence flow field

**3.3 Filtering in the modal space**

Based on the DMD analysis in section 3.2, for accelerating the convergence of the steady state flow, a certain number of snapshots are selected in the pseudo-time iterative steps, such that $A = [U_1, U_2, \cdots, U_N]$, where $U$ denotes each column vector consisted of the conservative flow variables for all the grids, and $N$ is the number of snapshots. According to the equation (20), the flow variables at any time step in modal space can be expressed in equation (21)

$$U_i = c_{i1}\Phi_1 + c_{i2}\Phi_2 + \cdots + c_{iN}\Phi_N \tag{21}$$

Because the DMD modes are sorted according to the different frequencies of the flow field, the equation (21) can be considered as the superposition of all the single-frequency components in modal space. Therefore, we can easily project the flow field solutions of the physical space to the modal space through the DMD analysis, and the different frequency components of the solution errors can be separated clearly. Different from the traditional multigrid method, we directly truncate all the higher-frequency modes in the modal space but only the first-order DMD mode is retained. And then, perform an inverse projection of the solutions to the physical space as the initial values for the next step. The proposed mode multigrid method can effectively avoid the complicated process of coarsening computational mesh since it is very convenient to truncate the high-frequency errors in the modal space. All the high-frequency perturbations can be quickly and completely filtered out, and the convergence for the flow solver is significantly accelerated.

There are mainly two reasons for the proposed MMG method to accelerate the convergence of the steady flow field. Firstly, the MMG method can achieve the similar effect of the traditional multigrid method. The traditional geometric multigrid method damps the solution errors by using the successive coarsening grids. While the MMG method projects the flow field from the physical space to the modal space, and truncates all the higher-frequency components of the solution errors directly in the modal space. Both of the methods can filter out the perturbations of the flow field during the iterations so as to accelerate the convergence. The implementation of the MMG method is much more convenient since it is not limited to the physical mesh compared with the geometric multigrid method. Secondly, the MMG method can quickly and accurately predict the stable steady mode for the flow field. The first-order mode obtained by the DMD represents the convergence solutions for the steady state flow, and this stable mode can be accurately extracted from the previous time steps. Therefore, the flow field after all the higher-order modes truncation can approach the steady state flow field much more precisely. The iterative values approximating to the convergence solutions are assigned to the initial flow field, which efficiently accelerate the convergence of the flow solver.

## 4. Numerical tests

### 4.1 Subsonic inviscid flow past a NACA0012 airfoil

The subsonic inviscid flow past a NACA0012 airfoil is firstly used to evaluate the performance of the developed MMG method. The computational mesh is shown in Figure 6, which consists of 7038 triangular elements and 200 boundary points on the airfoil surface. The free stream Mach number is $Ma = 0.63$ and the angle of attack is $\alpha = 2°$. We adopt the second-order finite volume method and the implicit symmetric Gauss-Seidel scheme for time-marching, and the CFL number is set as CFL=2. The total number of the iteration steps is $N = 22859$, and the computed lift coefficient is $C_l = 0.33014$.

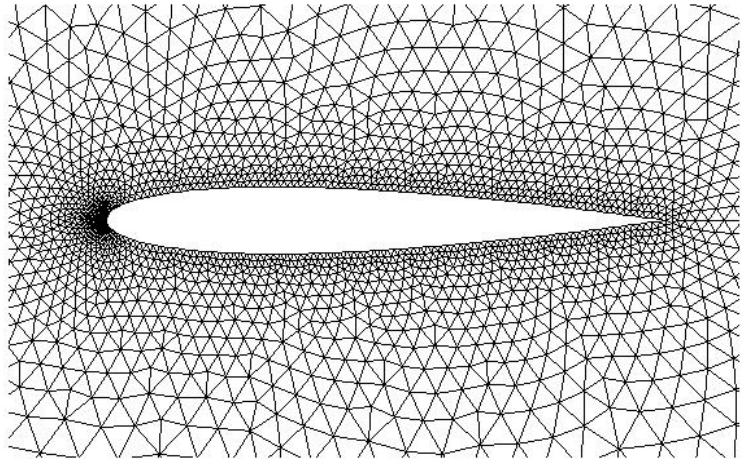

Figure 6 Grids near the NACA0012 airfoil surface for the inviscid subsonic flow

In order to verify the convergence acceleration technique, the proposed MMG method is used to re-compute the test case, where 40 snapshots at equal intervals are taken every 1000 iterative steps. The first 40, 10, 5 and 1 DMD modes are retained respectively, and the comparisons of the convergence histories are displayed in Figure 7, and the lift coefficients are shown in Figure 8. From the computational results, when retaining all of the 40 DMD modes (that is, no mode is truncated), the convergence history has almost no differences compared with the initial method. As the number of DMD modes decrease to 10 orders, only a slight acceleration effect can be achieved at the primary stage. And then the slope of the residual curve is nearly equal to the original iterative method. It illustrates that only a few parts of errors can be eliminated, but the flow field still contains high frequency perturbations, and the computational efficiency has not been improved evidently. When the DMD mode decreases to 5 orders, the acceleration convergence of the flow field is further improved, especially for the previous iterative steps. However, when the residual value drops to $10^{-9}$, the slope of the convergence curves is also equal to the original method, which is similar to the case of remaining 10 modes. When only the first DMD mode is preserved, that is, all the higher-frequency components are truncated, it only needs 6014 steps to converge. Compared with the original method, the number of iterative steps is reduced by 3.8 times, which can significantly speed up the convergence for the flow solver. Furthermore, we can also see form the Figure 8, the lift coefficient tends to the constant quickly, and the fluctuations of which are eliminated efficiently. The results demonstrate that the developed MMG method can efficiently reduce all the high-frequency components of the solution errors when all the high-order DMD modes are truncated in the modal space, and the convergence of the flow field is improved dramatically.

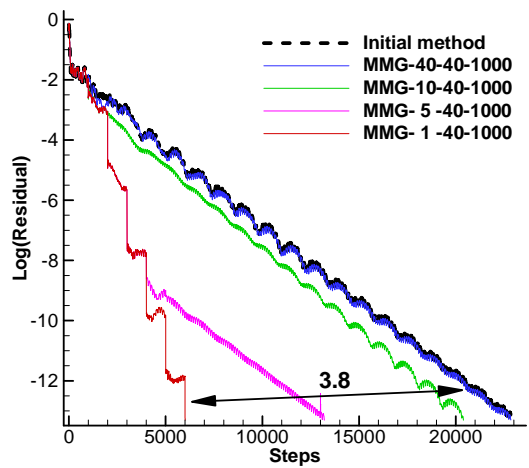 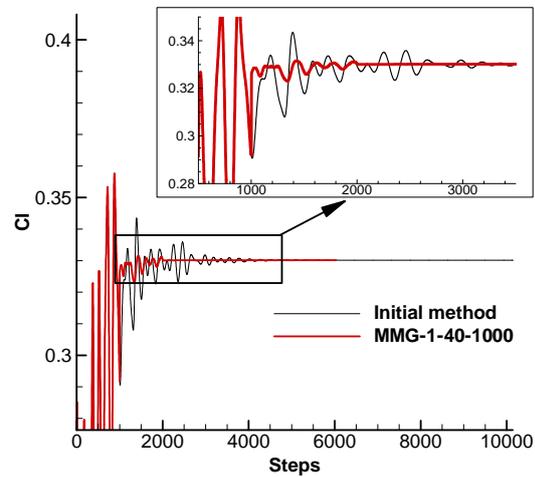

Figure 7 The convergence histories for different number of DMD modes

Figure 8 The response of lift coefficient for the initial method and the MMG method

From the convergence histories in Figure 7, it can be also seen that the residual value nearly drops 1 or 2 orders of magnitudes after each DMD analysis. Therefore, for the case of retaining only one mode, we compare the solution errors of the flow field before and after DMD analysis on each cell. Six times of DMD analysis are performed from the initial flow to the steady state flow, and the results of the second and sixth DMD analysis are shown in Figure 9, where the $x$-axis represents each cell number for all the grids, and the $y$-axis denotes the density errors between the current flow field and the convergence flow field. The results indicate that the absolute errors can be greatly reduced since the DMD analysis with modal truncation efficiently eliminates all the high-frequency perturbations for the flow field. The solutions obtained after the DMD analysis are much closer to the steady state flow. The results further verify that the first-order DMD mode can accurately represent the convergence solutions, and truncating the high-order DMD modes can obtain the flow distributions more approximating to the convergence solutions, which speeds up the convergence to steady state significantly.

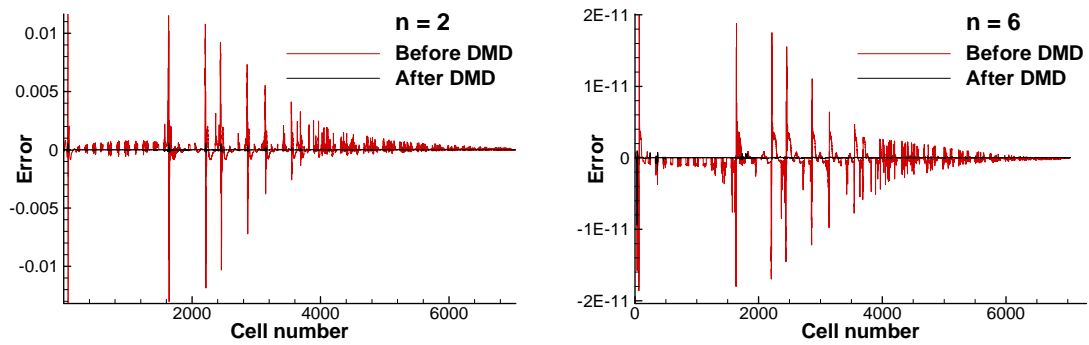

Figure 9 The comparisons of the solution errors before and after the DMD analysis

In order to further illustrate the effectiveness of the proposed MMG method, different numbers of snapshots are taken to compare the results. Also, only the first-order DMD mode is retained, and all of the higher-order modes are cut off completely. We take 40 snapshots at equal intervals, and implement the DMD analysis every 400, 800, 1000 and 1200 iterative steps respectively. The comparisons of the convergence histories are shown in Figure 10. Figure 11 displays the results for the case that 10, 20, 50 and 100 snapshots are taken every 1000 iterative steps. Compared with the original iterative method, the developed MMG method can effectively accelerate the convergence of the flow solver, and reduce the number of iterative steps robustly for all the different cases. Moreover, it can be found from the results, as a whole, the more the number of interval steps and the snapshots are taken, the better the acceleration convergence will be achieved. The effect of the MMG method is not very remarkable for the case of 400 interval steps or only 10 snapshots. The reason is that the flow information is inadequate to capture the steady stable mode. On the other hand, when the number of the interval steps is more than 1000 or the snapshots is more than 50, the results almost have no differences. It indicates that the number of flow field samples is sufficient enough to perform DMD analysis accurately. The detailed comparisons for the number of iterative steps and the computational time are shown in Table 1 and Table 2. Since the DMD analysis only requires a small amount of computational time compared with the original flow solver, the MMG method is much more efficient for applications. For this numerical case, the computational time can be reduced to 73.4% at most compared with the original iterative method.

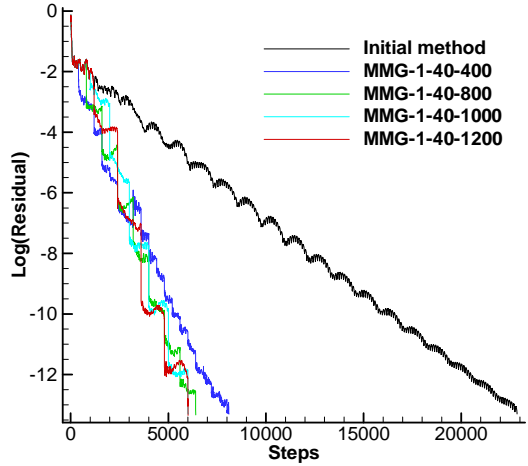 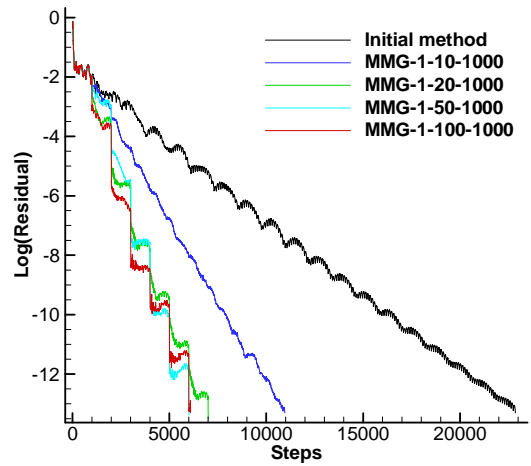

Figure 10 The convergence histories for different interval iterative steps

Figure 11 The convergence histories for different number of snapshots

Table 1 Reductions in number of iterations and CPU time for different interval steps of the inviscid NACA0012 airfoil

|                | Iterations | Reduction | CPU time | Reduction |
|----------------|------------|-----------|----------|-----------|
| Initial method | 22859      | —         | 362.8    | —         |
| MMG-1-40-400   | 8101       | 64.6%     | 132.5    | 63.5%     |
| MMG-1-40-800   | 6405       | 72.0%     | 103.1    | 71.6%     |
| MMG-1-40-1000  | 6014       | 73.7%     | 96.6     | 73.4%     |
| MMG-1-40-1200  | 6012       | 73.7%     | 96.4     | 73.4%     |

Table 2 Reductions in number of iterations and CPU time for different snapshots of the inviscid NACA0012 airfoil

|                | Iterations | Reduction | CPU time | Reduction |
|----------------|------------|-----------|----------|-----------|
| Initial method | 22859      | —         | 362.8    | —         |
| MMG-1-10-1000  | 10956      | 52.1%     | 174.5    | 51.9%     |
| MMG-1-20-1000  | 7003       | 69.4%     | 111.8    | 69.2%     |
| MMG-1-50-1000  | 6006       | 73.7%     | 96.8     | 73.3%     |
| MMG-1-100-1000 | 6095       | 73.3%     | 100.5    | 72.3%     |

We also compared the results of POD and DMD by using the developed MMG method, and 50 snapshots are taken every 1000 iterative steps. The convergence histories are displayed in Figure 12. Compared with the DMD, the POD does not significantly accelerate the convergence of the flow field. When only the first-order POD mode is retained, slight acceleration effect can be seen at the primary stage. And the slope of the residual curve is nearly equal to the original iterative method after the residual value drops to $10^{-5}$. When increasing the number of POD modes to five, the convergence will become worse. The results indicate that although the POD technique also has the effect of eliminating the high-frequency solution errors, however, the remained POD modes (even only the first mode) still contain different-frequency components of the errors. Furthermore, different from the DMD, the first-order POD mode is not the zero-frequency stable mode, which can not represent for the converged solutions of the steady state flow field. Therefore, compared with the DMD, the POD technique is not suitable for accelerating the steady state flow.

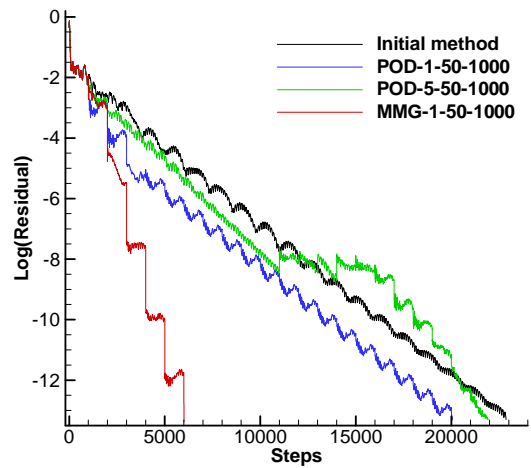

Figure 12 The convergence histories for the POD and DMD

Also, for this numerical test, we further increase the CFL number and perform the MMG method to compare the results. The CFL numbers are set as 10, 100 and 500 respectively. We take 50 snapshots every 800 iterative steps and only retain the first-order DMD mode. The comparisons of the convergence histories are shown in Figure 13, and the response of lift coefficients are displayed in Figure 14. It can be seen from the results that the convergence curves have little difference between the case of CFL=100 and CFL=500 for the primary iterative method. Therefore, it can be considered that there is no effect on the convergence rate when further increasing the CFL number. In the circumstance of the critical CFL number, the developed MMG method can still reduce the iterative steps 2~3 times less than the initial method, and the lift coefficients also converge quickly with little oscillation. The detailed number of iterative steps and the computational time are displayed in Table 3, which can reduce the computational time more than 50%.

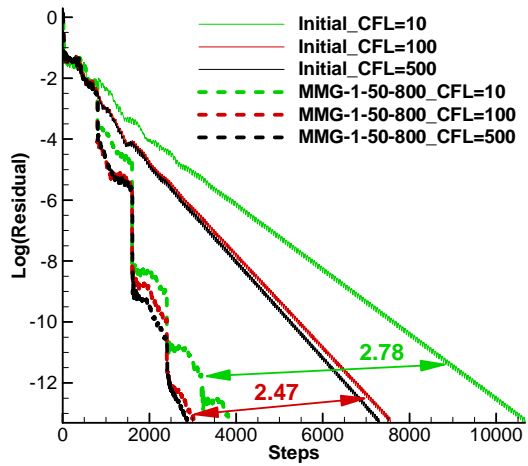

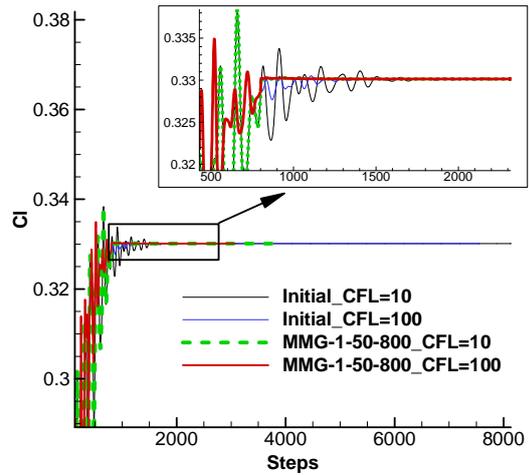

Figure 13 The convergence histories for different CFL numbers

Figure 14 The response of lift coefficient for different CFL numbers

Table 3 Reductions in number of iterations and CPU time for different CFL numbers of the inviscid NACA0012 airfoil

|                | Iterations | Reduction | CPU time | Reduction |
|----------------|------------|-----------|----------|-----------|
| Initial CFL=10   | 10665 | —     | 161.3 | —     |
| MMG CFL=10       | 3843  | 64.0% | 61.3  | 62.0% |
| Initial CFL=100  | 7563  | —     | 115.5 | —     |
| MMG CFL=100      | 3063  | 59.5% | 48.7  | 57.9% |
| Initial CFL=500  | 7308  | —     | 110.6 | —     |
| MMG CFL=500      | 2908  | 60.2% | 46.3  | 58.1% |

In order to verify the universality of the developed MMG method, we re-compute this numerical case by using the explicit third-order TVD Runge-Kutta method. The local time-step technique is added, and we fix the maximum CFL number as CFL=0.8. Similar to the implicit scheme, 40 snapshots are taken every 800, 1000, 1600 and 2000 iterative steps respectively. All the high-order modes are truncated, and only the first-order DMD mode is retained. The convergence histories and the lift coefficients are shown in Figure 15 and 16. We can clearly see from the results that the effect of acceleration is much more remarkable than the implicit schemes. The number of iterative steps is reduced by more than 6 times, and the lift coefficient approaches to convergence more quickly than the original method. The detailed number of iterative steps and the computational time are displayed in Table 4.

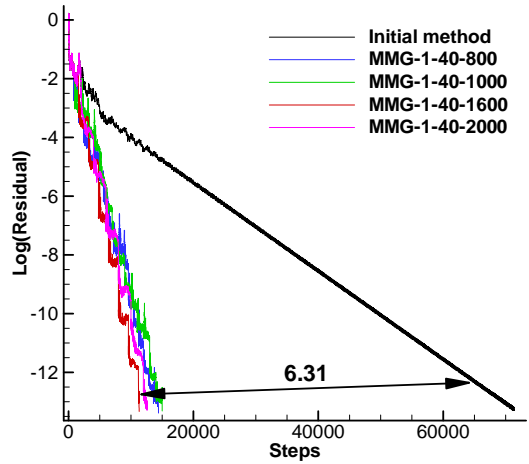
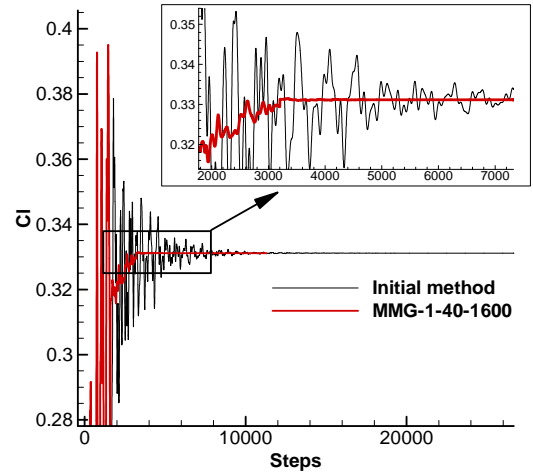

Figure 15 The convergence histories for the explicit Runge-Kutta scheme

Figure 16 The response of lift coefficient for Runge-Kutta scheme

Table 4 Reductions in number of iterations and CPU time for explicit Runge-Kutta time-marching method of the inviscid NACA0012 airfoil

|  | Iterations | Reduction | CPU time | Reduction |
| --- | --- | --- | --- | --- |
| Initial method | 71349 | — | 1219.2 | — |
| MMG-1-40-800 | 14404 | 79.8% | 248.9 | 79.6% |
| MMG-1-40-1000 | 15011 | 79.0% | 259.7 | 78.7% |
| MMG-1-40-1600 | 11308 | 84.1% | 195.5 | 84.0% |
| MMG-1-40-2000 | 12586 | 82.4% | 217.4 | 82.2% |

**4.2 Transonic flow past a RAE2822 airfoil**

In this section, we apply the proposed MMG method to the flow field with discontinuities. The test of transonic flow past a RAE2822 airfoil is used to validate the effectiveness of the method. The computational mesh is shown in Fig. 17, which consists of 5316 triangular elements and 200 boundary points on the airfoil surface. The free stream Mach number is $Ma = 0.75$ and the angle of attack is $\alpha = 3°$. In order to make a comprehensive survey for the method, the implicit and explicit time-marching methods, second-order and high-order accuracy schemes and different CFL numbers are performed to compare the results respectively. The high-order k-exact finite volume method and the accuracy preserved DDWENO limiter are used in this test. And the detailed introductions for the numerical method can be found in Reference [51]. The pressure isolines near the RAE2822 airfoil computed by the third-order scheme are displayed in Figure 18.

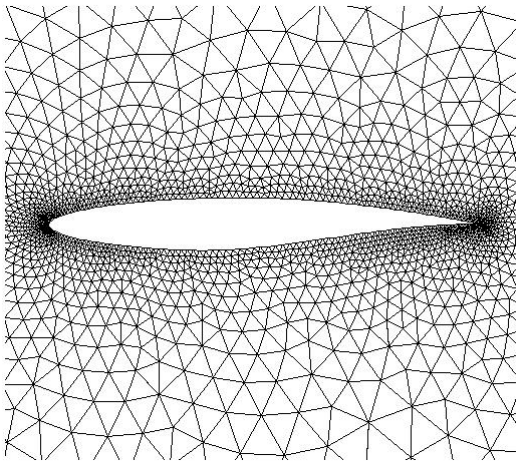 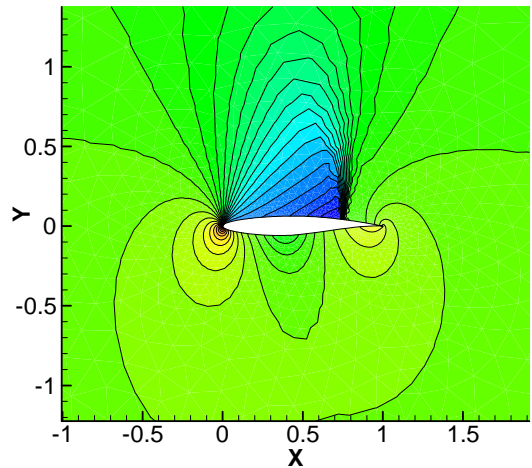

Fig. 17 The computational grids of RAE2822 airfoil

Figure 18 Pressure isolines near the RAE2822 airfoil

The MMG method is implemented to the second- and third-order schemes respectively. For the explicit time-marching scheme, the CFL number is set as 0.8, also the local time-step technique is used. And for the implicit time-marching scheme, we set the CFL number as 2, 100 and 500 respectively. The DMD analysis is performed by taking 50 snapshots every 1000 iterative steps. Figure 19 and 20 show the comparisons of the convergence histories for the second- and third-order schemes respectively. It can be seen from the results that the MMG method can still significantly accelerate the convergence for the discontinuous flow field. The number of iterative steps can be reduced by more than 3 times for the explicit scheme, and more than 2 times for the implicit scheme. For the high-order scheme, it also has good effect. As we can see that the number of iterative steps has almost no differences for the original method when the CFL numbers are 100 and 500. While in this circumstance, the MMG method can robustly reduce the number of iterative steps nearly 2 times. Figure 21 compares the pressure coefficients calculated by the second- and third-order schemes with the reference data in [52], and the entropy productions on the RAE2822 airfoil surface are displayed in Figure 22. The entropy distributions calculated by both the methods can completely coincide with each other, and the third-order scheme has much smaller entropy production than the second-order scheme. Therefore, the MMG method does not damage the numerical accuracy, and it is also independent on the spatial discrete schemes and the time-marching methods. In the case of the critical CFL number, the MMG method can still reduce the number of iterative steps, and significantly improve the efficiency of the flow solver. The detailed comparisons of the iterative steps and the computational time are shown in Table 5.

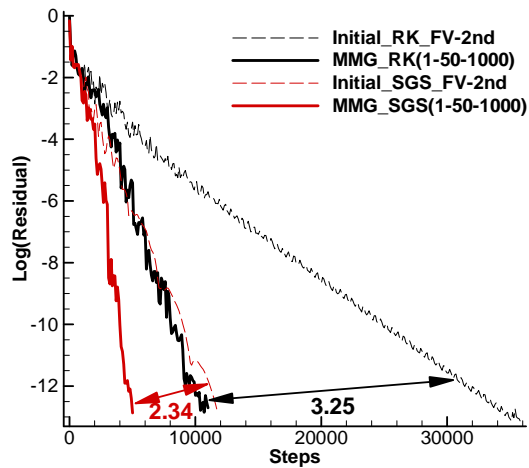

Figure 19 The convergence histories for RAE2822 airfoil computed by the second-order scheme

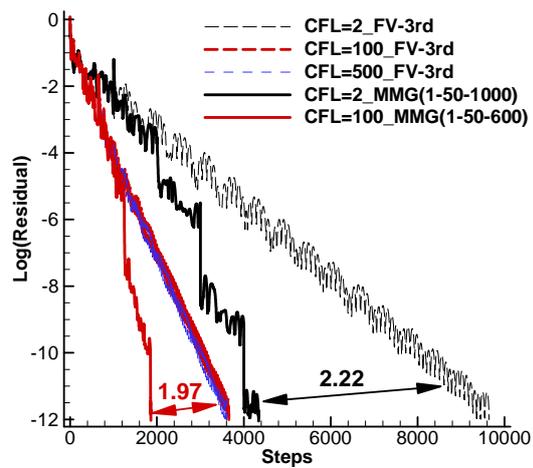

Figure 19 The convergence histories for RAE2822 airfoil computed by the third-order scheme

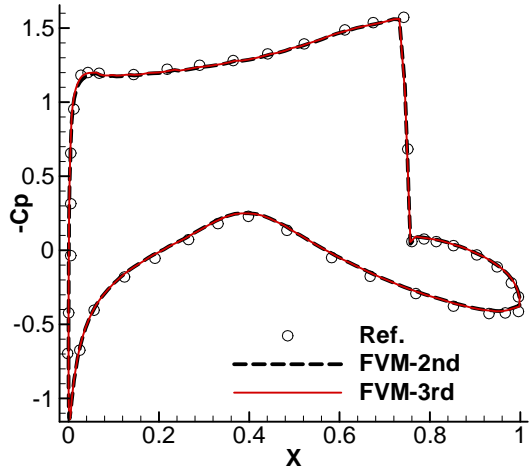 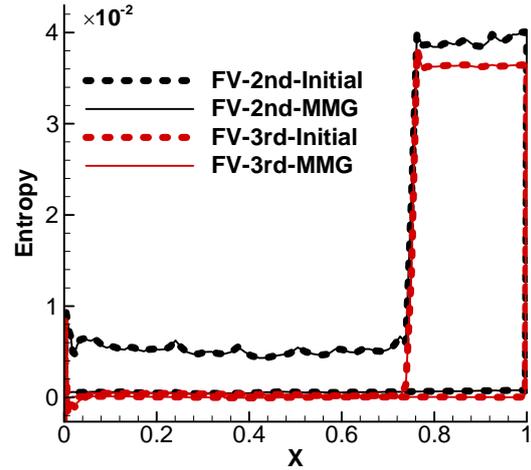

Figure 21 The pressure coefficients for the RAE2822 airfoil calculated by the second- and third-order schemes

Figure 22 The entropy production on the RAE2822 airfoil calculated by the initial method and the MMG method

Table 5 Reductions in number of iterations and CPU time of the transonic RAE2822 airfoil

|  |  | Iterations | Reduction | CPU time | Reduction |
|---|---|---|---|---|---|
| FV-2nd | Initial RK | 35831 | — | 904.5 | — |
|  | MMG RK | 11005 | 69.3% | 280.5 | 69.0% |
|  | Initial GS | 11713 | — | 177.2 | — |
|  | MMG GS | 5003 | 57.3% | 76.8 | 56.6% |
| FV-3rd | Initial CFL=2 | 9650 | — | 385.4 | — |
|  | MMG CFL=2 | 4343 | 55.0% | 174.6 | 54.7% |
|  | Initial CFL=100 | 3659 | — | 144.5 | — |
|  | MMG CFL=100 | 1861 | 49.1% | 74.2 | 48.7% |

### 4.3 Laminar flow past a NACA0012 airfoil

In this section, the developed MMG method is implemented for the viscous flow on anisotropic mesh. We consider a subsonic laminar flow past the NACA0012 airfoil at an angle of attack $\alpha = 0°$, the free stream Mach number $M_\infty = 0.5$, and the Reynolds number $Re_\infty = 5000$. The unstructured hybrid mesh employed for this calculation is depicted in Figure 23. It contains 18667 elements and 300 cells are distributed on the surface of the airfoil. We adopt the second- and third-order finite volume methods respectively, and the implicit symmetric Gauss-Seidel scheme for the time marching. The CFL number is set as CFL=2. The MMG method is performed by taking 50 snapshots every 1000 and 1500 iterative steps, and only the first DMD mode is retained. The convergence histories and the response of the drag coefficients are displayed in Figure 24 and 25 respectively. We can see from the results that the MMG method has no dependence on the mesh type, which is still suitable for the anisotropic hybrid meshes. The number of iterative steps can be reduced by 2 to 3 times both for the second- and third-order schemes. The oscillations of the drag coefficients are much slight than the original method. Figure 26 shows the streamlines of flow separation near the trailing edge computed by the third-order scheme. The comparisons of the drag coefficients on the airfoil with the reference data in [53] are displayed in Figure 27. The drag coefficients computed by the MMG method are completely consistent to the original method, which illustrates that the developed method can ensure the computational accuracy. The detailed comparisons of the iterative steps and the CPU time are shown in Table 6. The computational efficiency can be improved more than doubled.

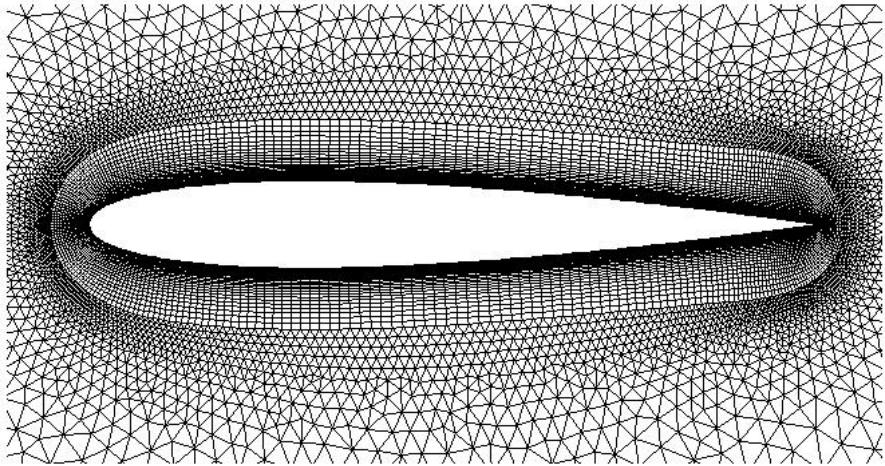

Fig. 23 Computational mesh for the laminar flow over the NACA0012 airfoil

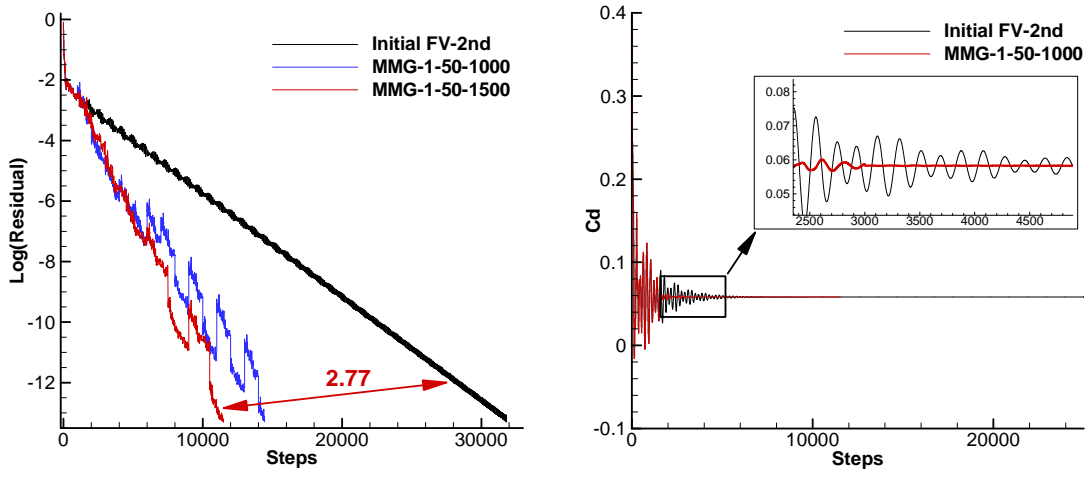

Figure 24 The convergence histories and the response of the drag coefficients for the laminar NACA0012 airfoil computed by the second-order scheme

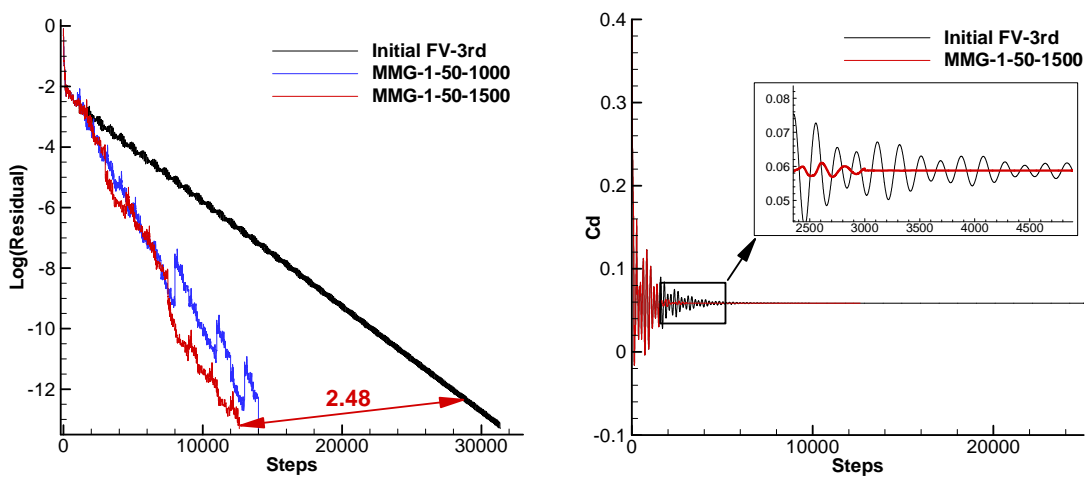

Figure 25 The convergence histories and the response of the drag coefficients for the laminar NACA0012 airfoil computed by the third-order scheme

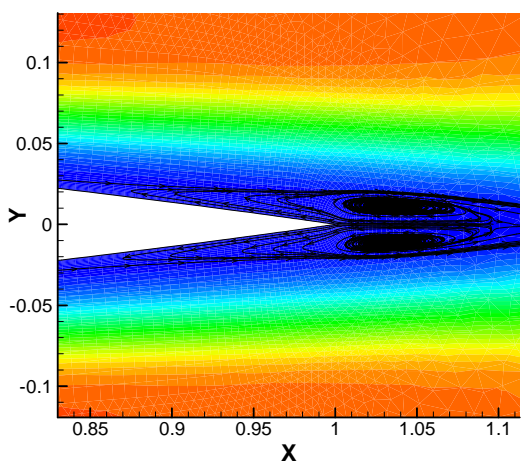
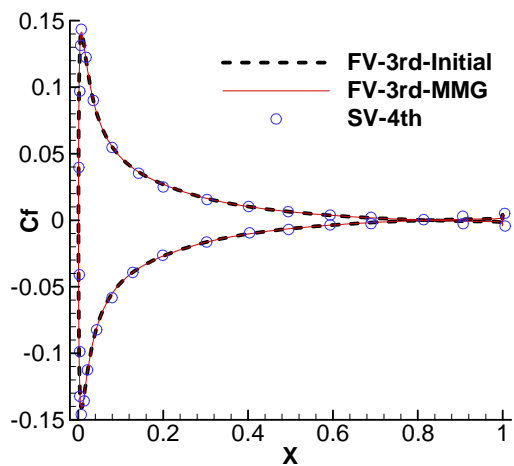

Fig. 26 Closed view of the small circulation bubble in the near wake for laminar flow over

Fig. 27 Comparisons of the computed skin friction coefficient over the NACA0012 airfoil

the NACA0012 airfoil

surface by the initial method and the MMG method with the reference data (fourth-order spectral volume method in Reference [53])

Table 5 Reductions in number of iterations and CPU time of the laminar NACA0012 airfoil

|  |  | Iterations | Reduction | CPU time | Reduction |
|---|---|---|---|---|---|
| FV-2nd | Initial method | 31823 | — | 1902.8 | — |
|  | DMD 1000-50-1 | 14443 | 54.6% | 875.2 | 54.0% |
|  | DMD 1500-50-1 | 11492 | 63.9% | 693.0 | 63.6% |
| FV-3rd | Initial method | 31358 | — | 3595.5 | — |
|  | DMD 1000-50-1 | 13017 | 58.8% | 1503.3 | 58.2% |
|  | DMD 1500-50-1 | 12219 | 61.0% | 1407.8 | 60.8% |

**4.4 Inviscid transonic flow over the ONERA M6 wing**

In this section, the three-dimensional ONERA M6 wing is used to verify the effectiveness of the MMG method. The ONERA M6 wing has a sweepback angle of $30°$, aspect ratio of 3.18 and taper ratio of 0.562. The airfoil section of the wing is the ONERA "D" airfoil, which has a 10% maximum thickness-to-chord ratio. The grid consists of 380188 arbitrary tetrahedral cells and 19522 boundary elements on the M6 wing. The surface mesh is displayed in Figure 28. We solve the Euler equations with the free stream Mach number Ma=0.84 and the angle of attack $\alpha = 3.06°$. The spatial discretization method and the time-marching method are the same as the ones in section 4.3. Figure 29 shows the pressure distributions on the upper surface of M6 wing.

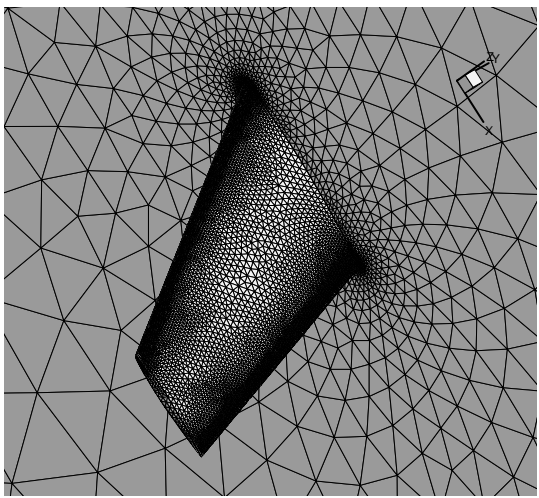 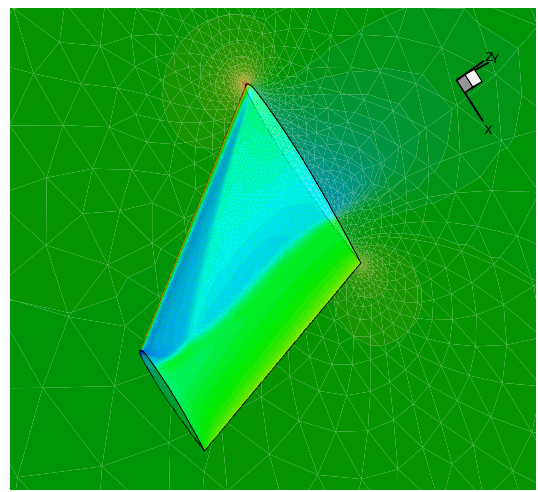

Fig. 28 The surface unstructured mesh of M6 wing

Fig. 29 The pressure distributions on the upper surface of M6 wing

The MMG method is performed by taking 20, 30 and 40 snapshots every 1000, 1500 and 2000 iterative steps respectively, and only the first DMD mode is retained. Figure 30 shows the comparisons of the convergence histories and the lift coefficient for the original method and the MMG method respectively. It can be seen from the results that the MMG method still has significant effectiveness for accelerating the convergence for the three-dimensional flow. The number of iterative steps can be reduced by nearly 5 times, and the response of the lift coefficient also converges much more quickly. Figure 31 compares the pressure coefficients in different spanwise stations with the reference data in [54]. The results calculated by the MMG method are completely the same as the original iterative method, which also well coincide with the reference data. The detailed comparisons of the iterative steps and the CPU time are shown in Table 7. We can see that there are little differences for selecting different number of snapshots, and the CPU time can be reduced to more than 75% compared with the original iterative method, which indicates that the MMG method can robustly accelerate the convergence for the flow solver. In addition, the MMG method needs to store some flow information in pseudo time iterations for DMD analysis. For this case, the sum of the volume cells is about 0.38 million, and 297MB, 446MB and 595MB of RAM is required for storing 20, 30 and 40 snapshots respectively by using double precision. That is, as an example for 1 million cells, only 150MB of RAM for a snapshots is required, which is easily affordable. Therefore, the proposed MMG method will not dramatically increase the computational memory, which can be also applied to accelerate the convergence of the flow field for the three-dimensional complex configurations.

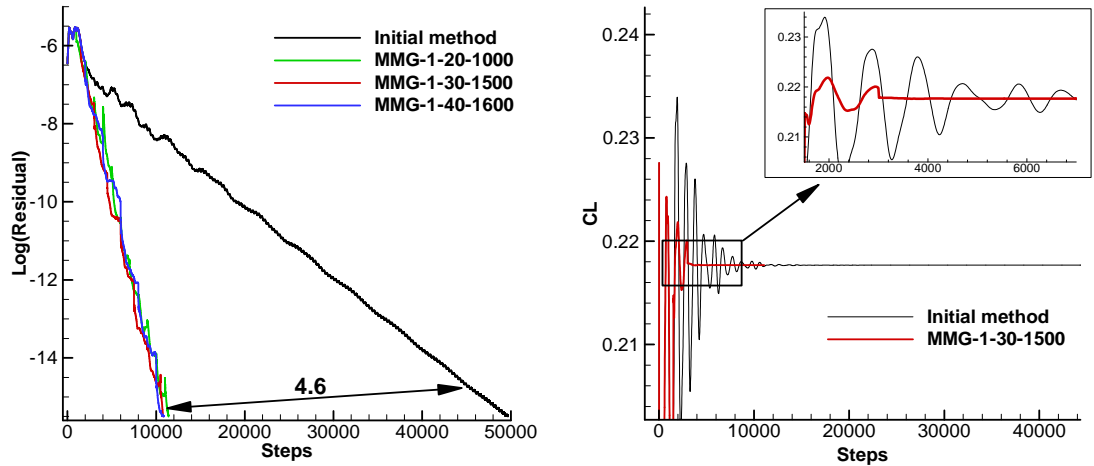

Figure 30 The convergence histories and the response of the lift coefficients for the M6 wing

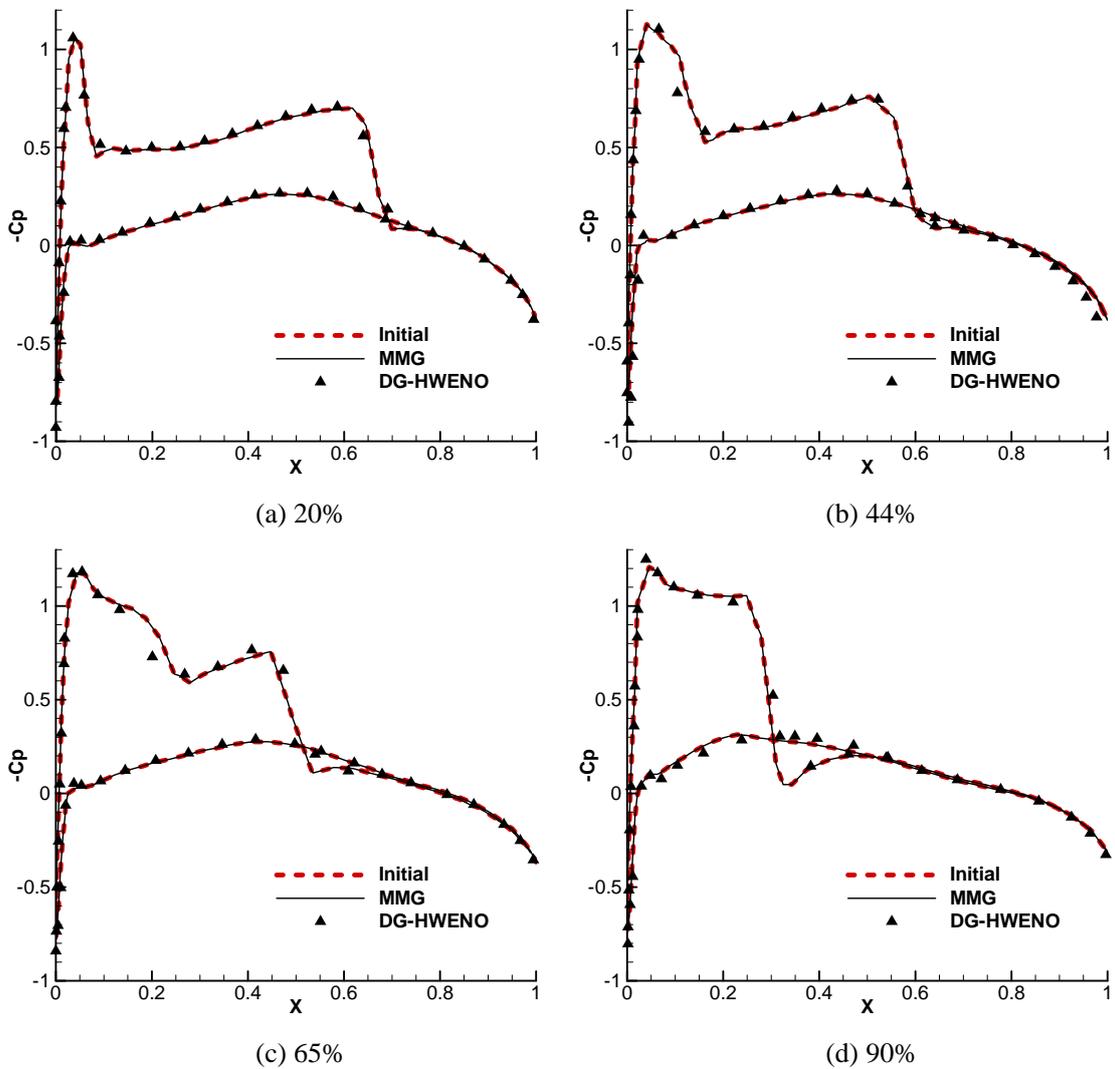

(a) 20%

(b) 44%

(c) 65%

(d) 90%

Fig. 31 The comparisons of pressure coefficient distributions for the wing section at different semi-span locations for the transonic M6 wing obtained by the initial method, MMG method and DG-HWENO in Reference [54]

Table 7 Reductions in number of iterations and CPU time of the M6 wing

|  | Iterations | Reduction | CPU time | Reduction |
|---|---|---|---|---|
| Initial method | 49670 | — | 61867.1 | — |
| MMG-1-20-1000 | 11416 | 77.0% | 14269.9 | 76.9% |
| MMG-1-30-1500 | 10797 | 78.3% | 13488.9 | 78.2% |
| MMG-1-40-2000 | 10866 | 78.1% | 13566.3 | 78.1% |

## 5. Conclusions

A novel mode multigrid method has been proposed in this paper and applied successfully to accelerate the convergence of steady state flow. Several typical numerical tests have been used to verify the effectiveness of the method. The following conclusions can be drawn:

(1) Compared with the traditional multigrid, the proposed MMG method projects the flow field solutions from the physical space to the modal space based on the DMD technique. Truncating the high-frequency components in modal space can achieve the same effect of the traditional multigrid method, which ingeniously avoids the complicated process of the grid coarsening and the data transfer between the fine grids and coarse grids. Furthermore, the MMG method is independent on the numerical method and only needs the flow field solutions in a few iterative steps. It is easy to be implemented and convenient for parallel computing, and can be inserted to any flow solver.

(2) A novel illustration for the convergence of the steady flow has been made in the perspective of modal space according to the DMD analysis, which is that the first-order flow mode converges to a steady state and all of other high-order modes are damped to zero. The DMD technique has the ability to predict the stable state for the oscillating flowfield solutions, and the first-order DMD mode is the zero-frequency stable mode, which can represent the convergence solution for the steady flow field. While, the POD technique can not accurately predict the stable mode, so it is not suitable for accelerating the convergence of the steady state flow.

(3) The developed MMG method requires only a small amount of memory and CPU time since it only needs to store a few snapshots for DMD analysis. Several numerical tests indicate that the MMG method can significantly improve the computational efficiency while ensuring the numerical accuracy. The iterative steps can be reduced by 3 to 6 times, and the CPU time can be reduced for 50%~80%.

(4) The developed MMG method has no dependence on computational mesh, and can be used in unstructured triangular grids and anisotropic hybrid grids both for viscous and discontinuous flows. Moreover, for the explicit/implicit time-marching method and second-/high-order schemes, the proposed method also has remarkable acceleration effect, which exhibits great potential in extensive applications on complex configurations and engineering.